\documentstyle[11pt,twoside]{article}

\begin{document}

\newcommand{\dd}{\,{\rm d}}
\newcommand{\ie}{{\it i.e.},\,}
\newcommand{\etal}{{\it et al.\ }}
\newcommand{\eg}{{\it e.g.},\,}
\newcommand{\cf}{{\it cf.\ }}
\newcommand{\vs}{{\it vs.\ }}
\newcommand{\zdot}{\makebox[0pt][l]{.}}
\newcommand{\up}[1]{\ifmmode^{\rm #1}\else$^{\rm #1}$\fi}
\newcommand{\dn}[1]{\ifmmode_{\rm #1}\else$_{\rm #1}$\fi}
\newcommand{\upd}{\up{d}}
\newcommand{\uph}{\up{h}}
\newcommand{\upm}{\up{m}}
\newcommand{\ups}{\up{s}}
\newcommand{\arcd}{\ifmmode^{\circ}\else$^{\circ}$\fi}
\newcommand{\arcm}{\ifmmode{'}\else$'$\fi}
\newcommand{\arcs}{\ifmmode{''}\else$''$\fi}
\newcommand{\MS}{{\rm M}\ifmmode_{\odot}\else$_{\odot}$\fi}
\newcommand{\RS}{{\rm R}\ifmmode_{\odot}\else$_{\odot}$\fi}
\newcommand{\LS}{{\rm L}\ifmmode_{\odot}\else$_{\odot}$\fi}

\newcommand{\Abstract}[2]{{\footnotesize\begin{center}ABSTRACT\end{center}
\vspace{1mm}\par#1\par
\noindent
{\bf Key words:~~}{\it #2}}}

\newcommand{\TabCap}[2]{\begin{center}\parbox[t]{#1}{\begin{center}
  \small {\spaceskip 2pt plus 1pt minus 1pt T a b l e}
  \refstepcounter{table}\thetable \\[2mm]
  \footnotesize #2 \end{center}}\end{center}}

\newcommand{\TableSep}[2]{\begin{table}[p]\vspace{#1}
\TabCap{#2}\end{table}}

\newcommand{\TableFont}{\footnotesize}
\newcommand{\TableFontIt}{\ttit}
\newcommand{\SetTableFont}[1]{\renewcommand{\TableFont}{#1}}

\newcommand{\MakeTable}[4]{\begin{table}[htb]\TabCap{#2}{#3}
  \begin{center} \TableFont \begin{tabular}{#1} #4 
  \end{tabular}\end{center}\end{table}}

\newcommand{\MakeTableSep}[4]{\begin{table}[p]\TabCap{#2}{#3}
  \begin{center} \TableFont \begin{tabular}{#1} #4 
  \end{tabular}\end{center}\end{table}}

\newenvironment{references}%
{
\footnotesize \frenchspacing
\renewcommand{\thesection}{}
\renewcommand{\in}{{\rm in }}
\renewcommand{\AA}{Astron.\ Astrophys.}
\newcommand{\AAS}{Astron.~Astrophys.~Suppl.~Ser.}
\newcommand{\ApJ}{Astrophys.\ J.}
\newcommand{\ApJS}{Astrophys.\ J.~Suppl.~Ser.}
\newcommand{\ApJL}{Astrophys.\ J.~Letters}
\newcommand{\AJ}{Astron.\ J.}
\newcommand{\IBVS}{IBVS}
\newcommand{\PASP}{P.A.S.P.}
\newcommand{\Acta}{Acta Astron.}
\newcommand{\MNRAS}{MNRAS}
\renewcommand{\and}{{\rm and }}
\section{{\rm REFERENCES}}
\sloppy \hyphenpenalty10000
\begin{list}{}{\leftmargin1cm\listparindent-1cm
\itemindent\listparindent\parsep0pt\itemsep0pt}}%
{\end{list}\vspace{2mm}}

\def\TYLDA{~}
\newlength{\DW}
\settowidth{\DW}{0}
\newcommand{\dw}{\hspace{\DW}}

\newcommand{\refitem}[5]{\item[]{#1} #2%
\def\REFARG{#3}\ifx\REFARG\TYLDA\else, {\it#3}\fi
\def\REFARG{#4}\ifx\REFARG\TYLDA\else, {\bf#4}\fi
\def\REFARG{#5}\ifx\REFARG\TYLDA\else, {#5}\fi.}

\newcommand{\Section}[1]{\section{#1}}
\newcommand{\Subsection}[1]{\subsection{#1}}
\newcommand{\Acknow}[1]{\par\vspace{5mm}{\bf Acknowledgements.} #1}
\pagestyle{myheadings}

\newfont{\bb}{timesbi at 12pt}

\def\thefootnote{\fnsymbol{footnote}}

\begin{center}
{\Large\bf Optical Gravitational Lensing Experiment.\\
Distance to the Magellanic Clouds\\ with the Red Clump Stars:\\
Are the Magellanic Clouds 15\% Closer\\
than Generally Accepted?\footnote{Based on 
observations obtained with the 1.3~m Warsaw telescope at the Las
Campanas  Observatory of the Carnegie Institution of Washington.}} 
\vskip1cm
{\bf A.~~U~d~a~l~s~k~i$^1$,~~M.~~S~z~y~m~a~{\'n}~s~k~i$^1$,~~
M.~~K~u~b~i~a~k$^1$,\\ 
G.~~P~i~e~t~r~z~y~\'n~s~k~i$^1$,~~ 
P.~~W~o~\'z~n~i~a~k$^2$,~~ and~~K.~~\.Z~e~b~r~u~\'n$^1$}
\vskip6mm
{$^1$Warsaw University Observatory, Al.~Ujazdowskie~4, 00-478~Warszawa, Poland\\
e-mail: (udalski,msz,mk,pietrzyn,zebrun)@sirius.astrouw.edu.pl\\
$^2$ Princeton University Observatory, Princeton, NJ 08544-1001, USA\\
e-mail: wozniak@astro.princeton.edu}
\end{center}
\vskip1cm
\Abstract{We present a new distance determination to the Large and Small 
Magellanic Clouds using the newly developed red clump stars method 
(Paczy\'nski and Stanek 1998). This new, single-step, Hipparcos
calibrated  method seems to be one of the most precise techniques of
distance determination  with very small statistical error due to large
number of red clump stars  usually available. 

The distances were determined independently along four lines-of-sight
located at opposite sides of each Magellanic Cloud. The results for each
line-of-sight are  very consistent. For the SMC we obtain the distance
modulus: $m-M=18.56\pm0.03\pm0.06$~mag (statistical and systematic
errors, respectively) and  for the LMC: ${m-M=18.08\pm0.03\pm0.12}$~mag
where systematic errors are mostly due to uncertainty in
reddening estimates. Both distances will be refined  and systematic
errors reduced when accurate reddening maps for our fields are
available.

Distance moduli to both Magellanic Clouds are ${\approx0.4}$~mag smaller
than  generally accepted values. The modulus to the LMC is in good
agreement with  the recent determinations from RR~Lyrae type stars and
upper limit resulting  from the SN1987A echo. We suspect that the
distance to the LMC and SMC is  shorter by about 15\% than previously
assumed: 42~kpc and 52~kpc, respectively. Calibrations of the
period-luminosity relation for Cepheids which give  overestimated
distances to the LMC and SMC are probably incorrect and require urgent
reanalysis. 

We also present our color-magnitude diagrams around the red clump for
the LMC  and SMC. We identify vertical red clump, first noted by
Zaritsky and Lin (1997),  in the color-magnitude diagram of both
Magellanic Clouds and we interpret it as an evolutionary feature rather
than unknown stellar population between the  LMC and our Galaxy.}{Magellanic 
Clouds -- Galaxies: distances and redshifts -- Hertzsprung-Russel (HR) 
diagram}

\Section{Introduction}
Two of our closest neighboring galaxies from the Local Group -- the
Magellanic  Clouds -- belong to the most important objects of the modern
astrophysics.  Hosting many objects commonly used as standard candles at
practically the same  distance they serve as ideal targets for
calibrating the distance scale in the  Universe. Therefore it is crucial
to have well established distances to both  Magellanic Clouds. 

Until recently, that is in the pre-Hipparcos era, it was generally
accepted  that the distance modulus to the Large Magellanic Cloud was in
the range of  ${m{-}M{=}(18.5{\div}18.6)}$~mag based mostly on the
calibration of the period-luminosity  (P--L) relation for Cepheids
(\eg Laney and Stobie 1994, Gieren \etal 1997 and  references
therein) and measurements of the supernova SN1987A echo (\cf 
Panagia \etal 1991, Sonneborn \etal 1997). In 1992 Walker
(1992) noted  ${\approx0.3}$~mag discrepancy between absolute magnitudes
of RR~Lyrae type stars from the Galaxy and the LMC at distance of
18.5~mag. Based on the new  statistical parallax solution for RR~Lyrae,
Layden \etal (1997) recalibrated  the absolute
magnitude-metallicity relation for these stars and found that the 
resulting distance modulus to the LMC might be significantly lower: ${m-
M=18.28 \pm0.13}$~mag. 

The latest determinations of the distance to the LMC come from Hipparcos
recalibration of the P--L relations for different type of
pulsating stars. Both Cepheid and Mira recalibrations lead to larger
distance moduli: Cepheids --  ${18.70\pm0.1}$~mag (Feast and Catchpole
1997) or ${18.57\pm0.11}$~mag (Madore and  Freedman 1998) and Miras --
${18.54\pm0.18}$~mag (van Leeuven \etal 1997). The value 
${m-M=18.50\pm0.10}$~mag has been adopted by the HST Extragalactic
Distance Scale Key  Project team (\eg Ra\-wson \etal 1997 and
other papers of the series) as the  distance to the LMC to which
extragalactic, Cepheid-based distance scale is  tied. 

On the other hand Hipparcos recalibration for RR~Lyrae based both on
direct  parallax determination and statistical parallaxes gives results
fully  confirming Layden \etal (1997) determination:
${18.26\pm0.15}$~mag (Fernley  \etal 1998). Similar results from
statistical parallaxes were recently  obtained by Popowski and Gould
(1998). Thus it seems very likely that the  RR~Lyrae stars give
${\approx0.25}$~mag shorter distance scale to the LMC.  Meanwhile,
recent reanalysis of the supernova 1987A echo by Gould and Uza (1998)
also suggests smaller distance modulus than previously derived ({\it
cf.}  Sonneborn \etal 1997, Lundquist and Sonneborn 1997) with
the upper limit as  low as ${m-M<18.37\pm0.04}$~mag. 

The distance to the Small Magellanic Cloud is poorly known. The distance
 modulus to the SMC is assumed to be about 18.9~mag (Westerlund 1990).
Massey  \etal (1995) obtained ${19.1\pm0.3}$~mag from
spectroscopic parallaxes. The most  recent determination based on the
Cepheid period-luminosity relation gives  similar value:
${m-M=18.94}$~mag (Laney and Stobie 1994). 

Summarizing this short review, the distance to the Large Magellanic Cloud
is at  least controversial and the generally accepted value
$m-M\approx18.5$~mag cannot  be now treated as established beyond any
doubt. The distance to the Small  Magellanic Cloud is assumed to be
${m-M\approx18.9}$~mag, still poorly known, and  therefore any new,
independent information about both distances is of the  greatest
importance. 

Recently Paczy\'nski and Stanek (1998) proposed a new, single step
method of  distance scale determination to the objects in the Galaxy and
neighboring  galaxies. The method bases on the fact that the red clump
giant stars seem to  be a very homogeneous group of objects and their
mean magnitude in the {\it I}-band is practically independent of their
color. Therefore the red clump  stars might be considered as excellent
standard candle candidates. The red  clump stars are very numerous
compared with other objects used as standard candles so far (typically
several orders of magnitude more numerous than \eg  pulsating
stars). Thus, their mean magnitudes can be obtained very precisely  with
small statistical error making the method very attractive for precise 
determination of distances in the Universe. 

The mean absolute magnitude of the red clump stars from the solar
vicinity was  determined by Paczy\'nski and Stanek (1998) who analyzed
luminosities of about  600 such stars with precise parallaxes (accuracy
better than 10\%) measured by  Hipparcos 
(${M^{\rm loc}_{I_0}=-0.185\pm0.016}$~mag). Paczy\'nski and Stanek 
(1998) applied the method for determination of the distance to the
center of  the Galaxy comparing $M^{\rm loc}_{I_0}$ with the mean 
{\it I}-band magnitude of  the red clump stars from the Baade's Window
obtained from photometry of the  first phase of the Optical
Gravitational Lensing Experiment (OGLE). The new method was also
applied by Stanek and Garnavich (1998) for determination of the 
distance to M31. Stanek and Garnavich (1998) also refined the maximum of
the  local red clump stars luminosity function by limiting the Hipparcos
sample of  the red clump stars to be volume rather than luminosity
limited and derived  ${M^{\rm loc}_{I_0}=-0.23\pm 0.03}$~mag from 228
red clump stars located  within the distance ${d<70}$~pc. Very large
number of red clump stars in the  Hipparcos sample makes the calibration
very sound comparing with  calibrations of other standard candles. It
should be noted that this  calibration might be even more precise in the
future when good {\it VI}  photometry is derived for more Hipparcos
stars. 

Although employing the red clump stars as distance indicator seems to be very  
attractive and can be possibly one of the most precise methods of distance  
determination one should be aware of potential systematic errors which can  
lead to errors in determined distance scale. The most important is proper  
determination of the interstellar extinction which can affect both target red  
clump luminosity as well as the local Hipparcos sample. However, the latter  
sample is very likely extinction free for ${d<70}$~pc or it is affected 
negligibly, \eg that one analyzed by Stanek and Garnavich (1998). Nevertheless 
precise determination of  extinction to the target stars is very important to 
avoid large systematic  errors. 

The second source of errors can be the differences in populations --
ages,  chemical composition etc., between the red clump stars of the
target object  and those in the solar neighborhood. This is, however, a
common problem of any  other method of distance determination when
similar objects from different  locations of the Universe are compared.
In case of the red clump stars  theoretical models predict that their
luminosity is very weakly dependent on  chemical composition and age
(Castellani, Chieffi and Straniero 1992, Jimenez,  Flynn and Kotoneva
1998). Nevertheless, the empirical fact of the very small  dispersion of
{\it I}-band magnitudes of the red clump stars as observed in the  solar
vicinity, Baade's Window and M31 (Paczy\'nski and Stanek 1998, Stanek 
and Garnavich 1998) should find theoretical explanation which could also
 answer the question about uncertainties introduced by comparison of
different  populations of the red clump stars. 

The natural consequence of the successful application of the red clump
method  for determination of distances to the Galactic center
(Paczy\'nski and Stanek  1998) and M31 (Stanek and Garnavich 1998) is a
distance determination to other  objects in the Local Group. The most
natural candidates are the Magellanic  Clouds, which are targets of the
microlensing surveys providing huge databases  of photometric data for
millions of stars in both Magellanic Clouds. 

The OGLE project is a long term microlensing survey which second phase
--  OGLE-II started at the beginning of 1997 (Udalski, Kubiak and
Szyma\'nski  1997). The Magellanic Clouds are the new targets of the
OGLE-II phase. Unlike  other microlensing surveys the OGLE project
observations are carried out with  the standard {\it BVI}-bands with
majority (${\approx75\%}$) observations  obtained in the {\it I}-band.
Thus, the data can be precisely transformed to the  standard {\it BVI}
system and applied directly to many projects unrelated to  microlensing.
In particular the OGLE observations are very well suited for the  above
mentioned red-clump method as the most of data is collected in the  
{\it I}-band. 

In this paper we apply the red clump stars method for determination of 
distances to both Large and Small Magellanic Clouds. Although OGLE-II
data  cover large fractions of both Magellanic Clouds we present here
distance  determination for two most west- and eastward lines-of-sight
toward both  Clouds where we assume extinction to be constant in the
first approximation.  Thus, the results should be considered as
preliminary and will be refined when  accurate maps of extinction based
on the OGLE-II data are derived. 

\Section{Observational Data}
All observations were collected with the 1.3-m Warsaw telescope at the
Las  Campanas Observatory, Chile, which is operated by the Carnegie
Institution of  Washington. The telescope and instrumentation as well as
the data pipeline has  been described by Udalski, Kubiak and Szyma\'nski
(1997). 

The photometric data for the SMC come from the just released "{\it BVI}
color  maps of the SMC" which will be publicly available after June 1,
1998 from the OGLE-II archive  (Udalski \etal 1998). The maps
provide three band photometry for eleven  driftscan strips covering
large part of the SMC (${\approx2.5}$ square  degree). For this study we
limited ourselves to the two most westward scans:  SMC$\_$SC1, and
SMC$\_$SC2 and two eastward scans: SMC$\_$SC10 and SMC$\_$SC11 
(Table~1). They are the least dense regions of the SMC, where we believe
the  extinction is small and constant. Each strip covers 
${\approx14\zdot\arcm2\times57\arcm}$ on the sky, and about
100~000--150~000  stars were detected in each of them. Photometric
reduction procedure is  described in detail in Udalski \etal
(1998). The magnitudes of stars are the average of about
100--115 and 20--25 observations in the $I$ and  {\it V}-bands,
respectively. Observations span the period from Jun.~26, 1997  through
Feb.~11, 1998. The data were reduced to the standard system based on 
observations of standard stars from selected Landolt's fields (Landolt
1992)  obtained on 20--22 and 5--8 nights for the {\it I} and {\it
V}-bands, respectively. The  accuracy of the zero point is better than
0.01~mag for both bands. Comparison  of the photometry with some
previous measurements showing very good agreement  can be found in
Udalski \etal (1998). 

\MakeTable{lcc}{12.5cm}{Coordinates of the SMC and LMC fields}
{
\hline
Field        &  RA$_{2000}$             & DEC$_{2000}$\\
\hline
SMC$\_$SC1   & $0\uph37\upm50\zdot\ups9$ & $-73\arcd29\arcm42$\arcs\\
SMC$\_$SC2   & $0\uph40\upm53\zdot\ups1$ & $-73\arcd17\arcm29$\arcs\\
SMC$\_$SC10  & $1\uph04\upm50\zdot\ups5$ & $-72\arcd24\arcm47$\arcs\\
SMC$\_$SC11  & $1\uph07\upm45\zdot\ups4$ & $-72\arcd39\arcm32$\arcs\\
&&\\
LMC$\_$SC15N & $5\uph01\upm18\zdot\ups0$ & $-68\arcd45\arcm52$\arcs\\
LMC$\_$SC14N & $5\uph03\upm47\zdot\ups3$ & $-68\arcd45\arcm39$\arcs\\
LMC$\_$SC19S & $5\uph43\upm47\zdot\ups1$ & $-70\arcd53\arcm33$\arcs\\
LMC$\_$SC20S & $5\uph46\upm16\zdot\ups9$ & $-71\arcd03\arcm40$\arcs\\
\hline
}

\vskip-12pt
Also for the LMC only the most east- and westward scans were selected: 
LMC$\_$SC20, LMC$\_$SC19, LMC$\_$SC14 and LMC$\_$SC15 (Table~1). All
these fields were  added to the main LMC targets of the OGLE-II search
in the 1997/98 observing  season and are still being extensively
observed. Full, three band photometry  of these fields will be released
in the future, when at least 100 observations  in the {\it I}-band and
30 in the $B$ and {\it V}-bands are collected. Mean magnitudes  used in
this paper were obtained from about 50 and 5 observations in the $I$ 
and {\it V}-bands, respectively. Thus, accuracy of individual stellar
magnitudes  is somewhat lower than that of the SMC data. Calibrations of
the data to the  standard system come from 9 and 2 nights only for the
$I$ and {\it V}-bands,  respectively. However, accuracy of the zero
point for the {\it I}-band is about  0.01~mag and 0.015~mag for the 
{\it V}-band. The data cover the period of Oct.~5,  1997 through Feb.~13,
1998. Selected fields are located at the edge of the LMC  bar and about
200~000 stars were detected and measured in each of them. 

We browsed through literature and compared our LMC photometry with already 
published, reliable CCD photometries. Unfortunately, dense, central  regions 
of the LMC where OGLE-II fields are located were very rarely observed  with 
the CCD technique, in particular in the {\it I}-band. We found only one  
reliable CCD photometry of regions overlapping with our fields: NGC~1850 (Sebo  
and Wood 1995). Unfortunately, Sebo and Wood (1995) do not provide their  
photometry of constant stars, only photometry of detected variable stars is 
available. Therefore, we extracted light curves of a few "stable" variable  
stars from our database, namely Cepheid variables, and compared our light  
curves with those of Sebo and Wood (1995). Fig.~1 presents sample 
{\it I}-band light curves of two Cepheids. We do not see any significant 
shifts in the  magnitude scale fully confirming that our zero points are 
determined  correctly. 

\Section{Red Clump in the Magellanic Clouds}
Figs.~2 and 3 present part of the $I$ {\it vs.} ${V-I}$ color-magnitude
diagram  (CMD) around the red clump for all SMC and LMC fields,
respectively. Because  of larger and non-uniform extinction in the LMC
fields we decided to analyze  only the least dense 1/3 part of the whole
driftscan strip of each field. We  designate such subfields with the
letter added in the field name describing location  of the subfield on
the strip: N -- northern part, S -- southern part. Size of  such trimmed
subfield is ${14\zdot\arcm2\times19\arcm}$ and equatorial  coordinates of
the center are given in Table~1. Figs.~2 and~3 indicate that at least
in the first approximation extinction is more or less constant in 
so selected fields, although in some fields, in particular LMC$\_$SC19S,
some non-uniformities of extinction are still present. Direction of the 
reddening is given by an arrow in each Figure. 

The color-magnitude diagrams of the SMC and LMC fields show all the features  
characteristic for CMDs of the Magellanic Clouds and resemble nicely synthetic 
CMD of the LMC obtained by Gallart (1998). Superimposed on the red giant  
branch, the red clump is very compact similarly as M31, Baade's Window and  
Hipparcos red clumps. There is no trace of the horizontal branch stars strip  
extending blueward from the red clump what differentiates Magellanic Clouds  
from M31 and suggests somewhat different stellar population content in these  
galaxies. 

Two additional features in the CMDs are clearly recognizable. First, the 
asymptotic giant branch bump brighter by about 1~mag and redder than the  red 
clump (Gallart 1998). The second feature is vertical extension of the red  
clump called vertical red clump (VRC) by Zaritsky and Lin (1997) and  
interpreted by them as a new, unknown stellar population located along the  
line-of-sight between the LMC and our Galaxy. However, it should be noted that 
this feature is not always present in our CMDs of the LMC fields (\eg 
LMC$\_$SC19S, but in those cases it could be smeared by differential 
extinction). It shows up in all SMC diagrams however, but there it bends a 
little blueward. It is also noticeable in the CMD of the Hipparcos local stars 
(\cf Paczy\'nski and Stanek 1998, Fig.~2). Therefore Zaritsky  and Lin (1997) 
interpretation seems to be incorrect -- the VRC is more  likely an 
evolutionary feature (\cf Beaulieu and Sackett 1998, Gallart 1998). 

\Section{Distance Determination}
To derive distances to the Magellanic Clouds, we followed the procedure 
described in Paczy\'nski and Stanek (1998) and Stanek and Garnavich
(1998). 

In the first step, we dereddened our CMDs. Unfortunately, the extinction
maps  for our fields derived from the OGLE-II photometry are not yet
available. Therefore, we decided to estimate extinction based on
extinction determination  from the literature. We limited ourselves to
papers with direct determination of  the reddening for the Magellanic
Clouds rather than to all-sky maps as  complexity of the Magellanic
Clouds extinction makes former determination more  reliable. As our fields
cover relatively large area on the sky (0.075 and 0.22  square degrees
for the LMC and SMC, respectively) which smooths reddening, we  decided
to use the mean extinction values adjusted to the general trends 
resulting from available extinction maps. 

In case of the SMC the situation is relatively simple. All determinations show  
that extinction toward the SMC, except for its central parts, is small. Our 
fields selected for distance determination are located outside the densest 
regions, therefore the reddening should be rather uniform which is confirmed 
by our CMDs. A direct extinction estimate toward our field SMC$\_$SC11 has 
been published. Mighell, Sarajedini and French (1998) derived ${E(B-
V)=0.08\pm0.03}$~mag toward the globular cluster NGC~416 which is located in 
this field. This is close to the mean reddening of the SMC given by Grieve and 
Madore (1986) or  Massey \etal (1995): ${E(B-V)=0.09}$~mag, and in general 
agreement with the extinction map of Grieve and Madore (1986). Therefore we 
adopted that value for both  eastward SMC fields -- SMC$\_$SC10 and 
SMC$\_$SC11. On the opposite side of  the SMC -- in the fields SMC$\_$SC1 and 
SMC$\_$SC2, there are no measurements  available. However this region is also 
far from the central bar and we can  safely adopt the same value for both 
westward fields. 

The situation is much more complicated in our LMC fields. Available maps of 
extinction show that reddening in the LMC is much more clumpy and the mean  
${E(B-V)}$ estimates range from 0.13 to 0.20~mag (\cf Harris, Zaritsky and  
Thompson 1997). The most recent maps (\eg Harris, Zaritsky and Thompson 1997,  
Oestreicher and Schmidt-Kaler 1996) show that extinction can vary  
significantly from place to place in the LMC. Our westward fields LMC$\_$SC14N  
and LMC$\_$SC15N are located relatively close to the Harris, Zaritsky and  
Thompson (1997) map which shows a trend of smaller extinction in the direction  
of our fields. Thus we adopt ${E(B-V)=0.17}$~mag for these fields, also in 
reasonable agreement with the map of Oestreicher and Schmidt-Kaler (1996). Our 
eastward fields are located about two degrees south  from the region of larger 
extinction (Oestreicher and Schmidt-Kaler 1996) and therefore we adopt there a 
little larger reddening: ${E(B-V)=0.20}$~mag.

As we already mentioned, the extinction uncertainty is the main source
of the  distance determination error. Therefore, to be on the safe side,
in both cases  we adopted relatively large error of reddening
${\sigma_{E(B-V)}=0.06}$~mag  which represents mostly uncertainty in
absolute extinction estimate rather  than differential extinction in the
field (compactness of the red clumps in the LMC, see below, suggests
small differential extinction). It should be  stressed here that
for more accurate distance determination, in particular to  the LMC, we
need much more precise reddening values which we hope to obtain  for our
fields from the OGLE-II photometric databases. Another possibility is  to
analyze the fields in the halo of the LMC, less affected by extinction
and  we plan to add such fields to the OGLE-II program in the future. 

Adopted reddening for all our fields with its error and
corresponding  extinction in the {\it I}-band as well as ${E(V-I)}$
(assuming standard extinction  curve -- ${A_I=1.96\times E(B-V)}$ and
${E(V-I)=1.28\times E(B-V)}$, Schlegel, Fink\-bei\-ner and Davis 1998)
are listed in Table~2. 

\MakeTable{lccc}{12.5cm}{Assumed reddening}
{
\hline
Field        & $E(B-V)$      &  $A_I$        & $E(V-I)$\\
             & [mag]         &  [mag]        & [mag]\\
\hline
SMC$\_$SC1   & $0.08\pm0.03$ & $0.16\pm0.06$ & $0.10\pm0.04$\\
SMC$\_$SC2   & $0.08\pm0.03$ & $0.16\pm0.06$ & $0.10\pm0.04$\\
SMC$\_$SC10  & $0.08\pm0.03$ & $0.16\pm0.06$ & $0.10\pm0.04$\\
SMC$\_$SC11  & $0.08\pm0.03$ & $0.16\pm0.06$ & $0.10\pm0.04$\\
&&&\\
LMC$\_$SC15N & $0.17\pm0.06$ & $0.33\pm0.12$ & $0.22\pm0.08$\\
LMC$\_$SC14N & $0.17\pm0.06$ & $0.33\pm0.12$ & $0.22\pm0.08$\\
LMC$\_$SC19S & $0.20\pm0.06$ & $0.39\pm0.12$ & $0.26\pm0.08$\\
LMC$\_$SC20S & $0.20\pm0.06$ & $0.39\pm0.12$ & $0.26\pm0.08$\\
\hline
}

In the second step, we selected the red clump stars in the magnitude
range  ${16.8<I_0<18.8}$~mag for the LMC and ${17.3<I_0<19.3}$~mag for
the SMC, and  color range ${0.8<(V-I)_0<0.95}$~mag. We limited the color
range from the blue  side to have exact overlap with the local Hipparcos
sample of the red clump  stars and from the red side to cut the tail of
more reddened stars and background red giant branch stars. Total number
of stars selected in these ranges for each  field is given in Table~3. 

Then, we prepared histograms of the magnitude distribution of the red clump 
stars with 0.02~mag bins and fitted a function given by Stanek and Garnavich 
(1998): 
$$n(I_0)=a+b(I_0-I_0^{\rm max})+c(I_0-I_0^{\rm max})^2+
\frac{N_{RC}}{\sigma_{RC}\sqrt{2\pi}}
\exp\left[-\frac{(I_0-I_0^{\rm max})^2}{2\sigma^2_{RC}}\right]\eqno(1)$$
which describes distribution of the red giant branch stars (three first
terms) with superimposed Gaussian distribution of the red clump stars.
Maxima of distribution and standard deviations $\sigma_{RC}$ are given
in Table~3. One should note small values of $\sigma_{RC}$
(0.13--0.16~mag for the LMC and 0.16--0.17~mag for the SMC) confirming
compactness of the red clumps in our fields. Figs.~4 and~5 present
histograms of the luminosity function of the red clump stars in our
fields with the best fit function given by Eq.~(1).

To check if our limitation of the ${(V-I)_0}$ color of the red clump
stars  does not affect derived maxima of luminosity function we repeated
calculations  including all red clump stars: ${(V-I)_0>0.7}$~mag. No
statistically significant  difference was found. 

Finally, we derived the distance moduli ${m-M}$ to our eight fields
assuming the mean absolute magnitude of the red clump stars that of the local
Hipparcos sample: ${M^{\rm loc}_{I_0}=-0.23\pm0.03}$~mag (Stanek
and Garnavich 1998). Results with corresponding errors are given in Table~3. 
We distinguish between errors from two sources: statistical -- resulting from 
uncertainties of the mean magnitudes of target and the local Hipparcos 
sample, and systematic -- resulting mostly from uncertainties of extinction 
determination. As can be noted for all four lines-of-sight to both Magellanic 
Clouds, which are located at different sections of these galaxies, results of 
distance determination are very consistent. For the SMC the  distance is 
determined with better accuracy. For the LMC the uncertainty is larger but it 
will certainly be improved in the future when better reddening information is 
available. Thus results should be considered as preliminary. Consistent 
results from the east- and westward locations of both Magellanic Clouds 
indicate that our extinction approximation was reasonable. 

\vskip-10pt
\MakeTable{lcccccc}{12.5cm}{Distance determination to the SMC and LMC}
{
\hline
Field &
Number & 
$I_0^{\rm max}$ & 
$\sigma_{RC}$ & 
$m-M$ &
$\sigma^{m-M}_{\rm stat}$ &
$\sigma^{m-M}_{\rm syst}$ \\
& of stars &&&&&\\
\hline
SMC$\_$SC1   & 13845 & $18.302\pm0.003$ & 0.16 & 18.53 & 0.03 & 0.06\\
SMC$\_$SC2   & 17852 & $18.326\pm0.003$ & 0.16 & 18.56 & 0.03 & 0.06\\
SMC$\_$SC10  & 12603 & $18.332\pm0.003$ & 0.17 & 18.56 & 0.03 & 0.06\\
SMC$\_$SC11  & 11370 & $18.342\pm0.003$ & 0.16 & 18.57 & 0.03 & 0.06\\
&&&&&&\\
LMC$\_$SC15N &  5600 & $17.860\pm0.003$ & 0.13 & 18.09 & 0.03 & 0.12\\
LMC$\_$SC14N &  6489 & $17.865\pm0.004$ & 0.13 & 18.10 & 0.03 & 0.12\\
LMC$\_$SC19S &  5641 & $17.871\pm0.004$ & 0.16 & 18.10 & 0.03 & 0.12\\
LMC$\_$SC20S &  4938 & $17.812\pm0.004$ & 0.15 & 18.04 & 0.03 & 0.12\\
\hline
}

\vskip-25pt
\Section{Discussion}
Results of distance determination to the Magellanic Clouds with the red
clump stars method (Table~3) lead to the surprising conclusion. All of
our determinations are consistently ${\approx0.4}$~mag smaller than
generally  accepted distance moduli to the Magellanic Clouds. 

There might be a few reasons of the discrepancy:

-- error of the zero point of the OGLE photometry. This is, however,
extremely  unlikely. In Section~2 we discussed accuracy of the OGLE
photometry and showed comparison with other reliable CCD photometries of
the Magellanic Clouds (\eg  Fig.~1). The OGLE mean magnitudes are
the average of tens of individual observations  and were tied with the
standard system based on observations collected on many  nights with
hundreds of observations of standard stars. We believe that our 
photometry constitutes a huge set of secondary {\it BVI} standards in
both  Magellanic Clouds and large systematic errors are excluded.

-- errors in reddening estimates for our fields. Certainly
overestimating of  the reddening leads to smaller distance moduli.
However, one should note that  even with the zero reddening the distance
moduli derived with the red clump method are  smaller than generally
accepted for the Magellanic Clouds (by at least  0.1~mag). But
extinction to the Magellanic Clouds does exist and for the LMC  the
foreground extinction is at least ${E(B-V)=0.075}$~mag,
${A_I=0.15}$~mag, and to the  SMC ${E(B-V)=0.037}$~mag,
${A_I=0.07}$~mag, (Schlegel, Finkbeiner and Davis 1998). Reddening  to
the SMC is much smaller and our distance determination is less affected
by  the reddening estimate. But consistently the distance modulus to the
SMC is  also smaller by ${\approx0.4}$~mag. 

-- the red clump stars method. We addressed this possibility in the 
Introduction. It seems unlikely from our present understanding of the
red  clump stars, but in principle, the mean {\it I}-band magnitude of
the red clump stars might  be a function of stellar population (although
it shows very small dispersion  in all objects investigated so far:
local sample, Galactic bulge, M31 and  Magellanic Clouds). We stress
here once again that this problem must be  carefully  investigated both
theoretically and observationally to answer  potential questions. 

As any of the above reasons does not seem to be satisfactory, we come to
the  very tempting conclusion that the distance determination to the
Magellanic  Clouds with the red clump method is correct, the distance
moduli are ${\approx  0.4}$~mag smaller than those generally accepted
and both Magellanic Clouds are  located about 15\% closer to our Galaxy
than previously assumed: the LMC at 42~kpc, and the SMC at 52~kpc. 

Our conclusion seems to be a strong argument in favor of the shorter distance 
to the LMC. Our determination is based on the independent, very 
straightforward, single-step method with minimum possibilities for the 
potential error sources. The new technique is calibrated with Hipparcos data 
with much better accuracy than calibration of any other standard candle 
(orders of magnitude more stars, very close objects with precise parallaxes). 
Independent determinations for four lines-of-sight located at different sides 
of each Magellanic Cloud yield very consistent results which makes them very 
sound. If we underestimated the reddening, the Magellanic Clouds are even 
closer to us. On the other hand if we assume only minimum, foreground  
reddening, which is certainly an underestimate, ($E(B-V)=0.075\pm0.015$~mag 
for the LMC and $E(B-V)=0.037\pm0.015$~mag for the  SMC, Schlegel, Finkbeiner 
and Davis 1998) the upper limits for distance moduli  are: $m-
M<18.29\pm0.03\pm0.03$~mag for the LMC and $m-M<18.65\pm0.03\pm0.03$~mag for 
the SMC. Still a few sigma gap between distance moduli resulting from the red 
clump method and presently accepted values  remains. 

Brocato \etal (1996) presented the CCD photometry of selected
globular  clusters in the LMC. We find there that one of the CMDs,
showing the region around  NGC~1786, contains a neat red clump.
Unfortunately Brocato \etal (1996) do not  provide their
photometry, but even visual inspection of their Fig.~1 indicates  that
the mean magnitude of the red clump stars is ${I=18.2\pm0.1}$~mag. 
NGC~1786 is located only one degree north from our field LMC$\_$SC15N so
it is  comparably affected by extinction (it falls almost into the
extinction map of  Harris, Zaritsky and Thompson 1997). Thus resulting
distance modulus from  NGC~1786 field and independent photometry is in
perfect agreement with our red  clump determinations in Table~3. 

As we mentioned in the Introduction, RR~Lyrae-based distance
determination  also seems to favor smaller distance modulus to the LMC.
All recent  determinations from statistical parallaxes and Hipparcos
calibrations of  RR~Lyrae stars are within one sigma agreement with our
determination. Also the  upper limit for the distance to the LMC from
the SN1987A light echo by Gould and Uza (1998) is compatible with our
determination. 

The only arguments for the larger distance to the LMC come then from the 
Cepheid P--L calibration (and the Mira variables but because of much lower  
accuracy this determination is much less reliable). However, because of its  
complex nature, the calibration might be simply incorrect. An incorrect 
account of possible population effects, mainly metallicity, could result in 
overestimated distances obtained with that method (\cf Sekiguchi and Fukugita 
1998, Sasselov \etal 1997). Our SMC distance determination also confirms this  
thesis: our distance modulus to the SMC is ${\approx0.4}$~mag smaller than  
that derived from Cepheid P--L (Laney and Stobie 1994). It should be also noted 
that there exists ${\approx0.3}$~mag discrepancy between Hipparcos calibrated 
 Cepheid P--L distance to M31 (Feast and Catchpole 1997) and recent 
determination with the red clump stars (Stanek and Garnavich 1998) and 
globular clusters (Holland 1998). Thus we would like to stress at this
point  the necessity of urgent reanalysis of the Cepheid P--L relation.
Very  precise, {\it BVI} light curves of hundreds of Cepheids from both
Magellanic  Clouds will be provided by the OGLE project to the
astronomical community later during this year and can be of great importance 
for such a project. 

Concluding, it seems very likely that Cepheid P--L relation gives 
distance moduli overestimated by about ${\approx0.4}$~mag (at least toward the 
Magellanic Clouds)  with all astrophysical consequences -- distances beyond 
M31, up to several Mpc  rely mainly on Cepheids and the LMC distance. Our 
independent distance measurement to the LMC  confirms the recent calibrations 
of the absolute magnitude--metallicity relation for RR~Lyrae stars giving 
fainter absolute magnitudes contrary to results of Chaboyer \etal (1998) and 
therefore leading to older ages of globular clusters  (Layden \etal 1996). On 
the other hand shorter distance to the LMC means  shorter distance to 
Cepheid-based galaxies and in turn larger Hubble constant  and shorter age of 
the Universe. The "age problem" gap increases. 

We believe that there exists at least one crucial test which can verify
our  present findings. Paczy\'nski (1997) proposed detached eclipsing
binaries as  an excellent method of distance determination. Having
precise light curve and  spectroscopic data one can determine the
distance to the eclipsing, well  detached system with accuracy of a few
percent. The catalog of eclipsing stars  in the Magellanic Clouds
suitable for such a distance determination with three  color light
curves, periods etc.\ will be released by the OGLE project in a few 
months. Faintness of the Magellanic Cloud stars will require the
largest, 10-m  class telescopes to collect good quality spectroscopic
data and a lot of  telescope time. However, importance of the problem
makes such a project the one of the highest priority and we hope it will
be undertaken soon. 

\Acknow{We would like to thank Prof.\ Bohdan Paczy\'nski for encouraging 
discussions and help at all stages of the OGLE project. We are indebted to Dr 
Krzysztof Stanek for useful coments and remarks. The paper  was partly 
supported by the Polish KBN grant 2P03D00814 to A.\ Udalski.  Partial support 
for the OGLE project was provided with the NSF grant  AST-9530478 to B.\ 
Paczy\'nski.} 

\bigskip

{\bf Note:} Based on theoretical isochrones fitting to the main 
sequence and the red clump in four fields in the LMC, Beaulieu and Sackett 
(1998, revised version of their paper) found significantly better fit assuming 
smaller distance modulus to the LMC (${m-M=18.3}$) providing thus, 
additional argument in favor of the shorter distance to the LMC.

\vspace{3cm}

\centerline{\bf Figure captions}

\vspace{1cm}

Fig.~1. Comparison of light curves of two Cepheids from NGC1850 field  
obtained by Sebo and Wood (1995) and OGLE-II project.

Fig.~2. Observed color-magnitude diagrams around the red clump for four 
fields in the SMC selected for distance determination. Arrow indicates 
reddening direction.

Fig.~3. Observed color-magnitude diagrams around the red clump for four 
fields in the LMC selected for distance determination. Arrow indicates 
reddening direction.

Fig.~4. Luminosity function of the red clump stars for four fields in the SMC. 
Bins are 0.02 mag wide. Solid line is the best fit obtained with the function 
given by Eq.~(1).

Fig.~5. Luminosity function of the red clump stars for four fields in the LMC. 
Bins are 0.02 mag wide. Solid line is the best fit obtained with the function 
given by Eq.~(1).
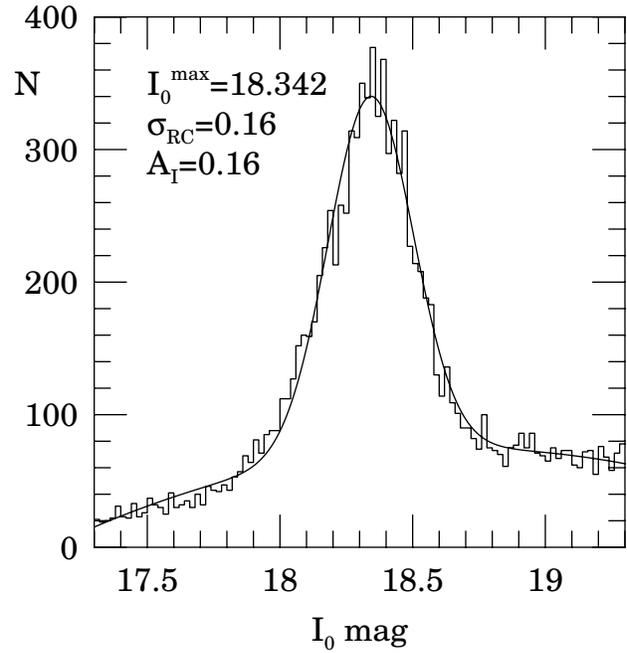
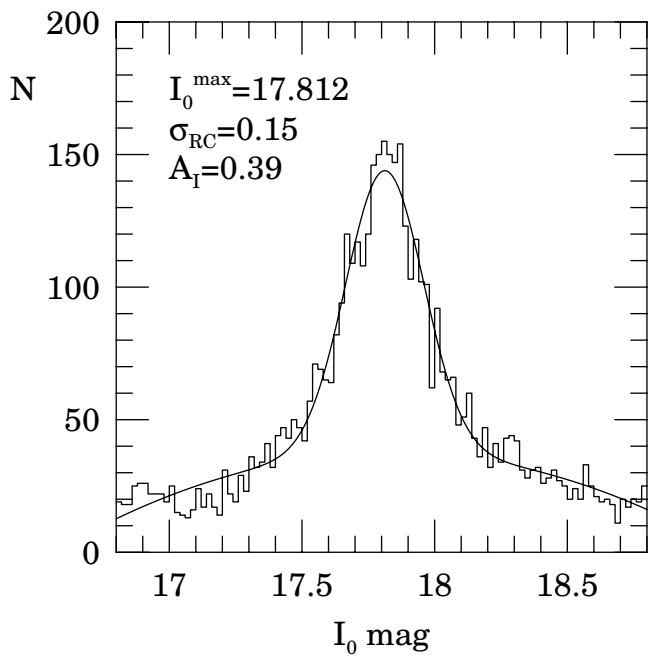
\end{document}